\documentclass[a4paper,11pt]{article}
\pdfoutput=1 % if your are submitting a pdflatex (i.e. if you have
\usepackage{jheppub} % for details on the use of the package, please
\usepackage{fancyhdr}
\usepackage{graphicx}
\usepackage[usenames,dvipsnames,svgnames,table]{xcolor}
\usepackage{xspace}
\usepackage{rotating}

\usepackage{braket}

\DeclareSymbolFont{extraup}{U}{zavm}{m}{n}
\DeclareMathSymbol{\varheart}{\mathalpha}{extraup}{86}
\DeclareMathSymbol{\vardiamond}{\mathalpha}{extraup}{87}

\usepackage{amsfonts}
\usepackage{pifont}

\newcommand{\be}{\begin{equation}}
\newcommand{\ee}{\end{equation}}
\newcommand{\bea}{\begin{eqnarray}}

\newcommand{\eea}{\end{eqnarray}}

\newcommand{\Rmnum}[1]{\expandafter\@slowromancap\romannumeral #1@}

\def\t{{\rm t}}
\def\f{{\rm f}}
\begin{document} 

%%%%%%%%%%%%%%%%%%%%%%%%%%%%%%%%%%%%%%%%%%%%%%%%
%%%%%%%%%%%%%%%%%%%%%%%%%%%%%%%%%%%%%%%%%%%%%%%%
%  Title, authors and abstract
%%%%%%%%%%%%%%%%%%%%%%%%%%%%%%%%%%%%%%%%%%%%%%%%
 \title{\boldmath The impact of non-minimally coupled gravity on vacuum stability}
 
%%%%%%%%%%%%%  Authors  Alphabetically %%%%%%%%%%%%%%%%%%%%
\author[1]{Olga Czerwi\'nska}
\author[1]{Zygmunt Lalak}
\author[1]{Marek Lewicki}
\author[1]{Pawe{\l } Olszewski}

%%%%%%%%%%%%%  Affiliations %%%%%%%%%%%%%%%%%%%%%%%%%%%%%
\affiliation[1]{
Institute of Theoretical Physics, Faculty of Physics, University of Warsaw ul. Pasteura 5, 02-093 Warsaw, Poland}
%%%%%%%%%  Email Addresses %%%%%%%%%%%%%%%%%%%
\emailAdd{zygmunt.lalak@fuw.edu.pl}
\emailAdd{olga.czerwinska@fuw.edu.pl}
\emailAdd{marek.lewicki@fuw.edu.pl}
\emailAdd{pawel.olszewski@fuw.edu.pl}
%%%%%%%%%%%%%%%%%%%%%%%%%%%%%%%%%%%%%%%%%%%%
%  Abstract

\abstract{
We consider vacuum decay in the presence of a non-minimal coupling  to gravity. 
We extend the usual thin-wall solution to include the non-minimal coupling.
We also perform a full numerical study and discuss the validity of the new thin-wall approximation.
Implications of a large cosmological constant, whose influence on the geometry boosts the tunneling rate, are discussed.
Our results show that the influence of the non-minimal coupling differs significantly between the  cases of Minkowski and  deSitter backgrounds.
In the latter the decay probability quickly decreases when the coupling grows and in fact the vacuum can be made absolutely stable simply due to introduction of the non-minimal coupling.
In the case of Minkowski background the effect is much weaker and the decay rate even increases for small values of the non-minimal coupling.

}


\maketitle
\flushbottom

%%%%%%%%%%%%%%%%%%%%%%%%%%%%%%%%%%%%%%%%%%%%%%%%%%%%%%%%%%%%%%%%%%%%%%%

\section{Introduction}
%\label{sec:intro}

The discovery of the Higgs boson gave rise to the important question of vacuum stability in the Standard Model.
Observed data indicate that the electroweak minimum in the SM effective potential is metastable, so the potential has a second minimum to which the electroweak vacuum may decay.
It has been of great interest to investigate features of such a process under more or less obvious modifications: addition of higher-dimensional interactions \cite{Lalak:2015usa}, approximate scale-invariance \cite{DiLuzio:2015iua}, the issue of gauge dependence \cite{Lalak:2016zlv}, \cite{Plascencia:2015pga}, the relation to primordial black holes \cite{Burda:2016mou}, to name a few.
The study of the gravitational impact on the metastability has been mostly following the classic work of Coleman and De Luccia \cite{Coleman:1980aw}.
%Using thin-wall approximation Coleman and De Luccia  showed that the decay of Minkowski vacuum into anti-deSitter one in gravitational background is heavily suppressed.
More recently a new study has been devoted to the validity of the thin-wall approximation in case of gravitational background \cite{Masoumi:2016pqb}.
Moreover, the influence of the additional scalar along with the curved spacetime in the gauge-less top-Higgs model has been investigated showing that the potential is modified both in the region of the electroweak minimum and in the region of large field strength, see \cite{Czerwinska:2015pch}. Quantum gravity corrections to the SM effective potential and their impact on vacuum stability have also been considered in \cite{Loebbert:2015eea}.

The question of the role of the non-minimal coupling $\xi$, between the scalar field and scalar curvature, in the process of vacuum decay is the central point of this note.
This coupling is required for the renormalizability of the scalar field in curved spacetime, even though it might be zero at a certain energy scale, and it is a crucial feature of the Higgs inflation model that is still allowed by the experimental data \cite{Ade:2015lrj}, \cite{Bezrukov:2007ep}. So far impact of the non-minimal coupling has been investigated in case of the inflationary background \cite{Herranen:2014cua}, \cite{Herranen:2015ima} and in the Standard Model case \cite{Isidori:2007vm}, \cite{Rajantie:2016hkj}.
In this paper we limit ourselves to the theories with a single scalar field and a renormalizable potential.
The seemingly simplified approach is dictated by the need to accommodate in a readable manner a wide spectrum of parameters all of which are controlling the influence of gravity. We vary not only the $\xi$. Tunneling both close and far from the thin-wall regime is discussed. We consider flat as well as closed (dS) geometry of the false vacuum, and closed (dS) or open (AdS) geometry of the true vacuum. Our qualitative discussion aims to be universally applicable in the plethora of contexts evoking the quantum tunneling in the presence of gravity. Examples could range from preventing catastrophes in phenomenological theories, through modeling past cosmological events like inflation \cite{DiVita:2015bha}, \cite{Bezrukov:2014ipa} and baryogenesis \cite{Lewicki:2016efe}, up to studies of the string theory landscape \cite{Demetrian:2005sr}.

The outline of the paper is as follows. In Section~\ref{sec:model} we present our  Lagrangian describing scalar field non-minimally coupled to gravity. In Section~\ref{sec:tunneling} we introduce the non-minimal coupling to the usual thin-wall approximation and calculate the appropriate numerical action for the bounce solution that supports our analytical approximation.
There we discuss the results exemplifying different regimes and investigate the influence of a large cosmological constant on the decay of the false vacuum. The connection between tunnelling via bubble nucleation and the Hawking-Moss instanton is also explored.

\section{The Model}\label{sec:model}
The main goal of this paper is to discuss the impact of the gravity on the vacuum decay process. On the particle physics side we consider a toy model describing a single neutral scalar field.
Standard gravitational interaction is supplemented by adding the non-minimal coupling of the scalar field to the Ricci scalar.
The Lagrangian takes the form
\begin{equation}
\label{eq:lagrngian}
\mathcal{L}= 
\frac{1}{2}(\partial \phi )^2-V+\frac{1}{2}\frac{R}{\kappa}\left(1-\xi \kappa  \phi^2 \right)
\end{equation}
with
\begin{equation}
\label{eq:potential}
V= -\frac{1}{4} a^2 (3 b - 1) \phi^2 + \frac{1}{2} a (b - 1) \phi^3 + \frac{1}{4} \phi^4+a^4 c.
\end{equation}
The potential is intentionally chosen to be very simple but at the same time informative as it exhibits all features we require to discuss tunnelling.

It has two minima: at $\phi=0$ and $\phi=a$. We will always consider a scenario when the field is initially in a homogeneous configuration in the more shallow minimum (or {\it false} vacuum) at $\phi=0$ which we will denote by $\phi_{\rm f}$.
And we consider tunnelling to the second deeper minimum (or {\it true} vacuum) at $\phi_\t=a$.

We use natural units where $M_p=1$, and, in our considerations, if not explicitly stated otherwise, we always set $a=1$, which means that the true vacuum is positioned at the Planck scale. Since $a$ is the only dimensionfull parameter, decreasing it simply corresponds to pushing the Planck scale further away, and decreasing the gravitational effects, bringing our results closer to flat spacetime case.

The constant $c$ is responsible for the character of our initial false vacuum and in this paper we focus on a de Sitter false vacuum case which means $c>0$ and Minkowski false vacuum with $c=0$. 

\begin{figure}[ht]
\begin{center}
\includegraphics[width=0.65\textwidth]{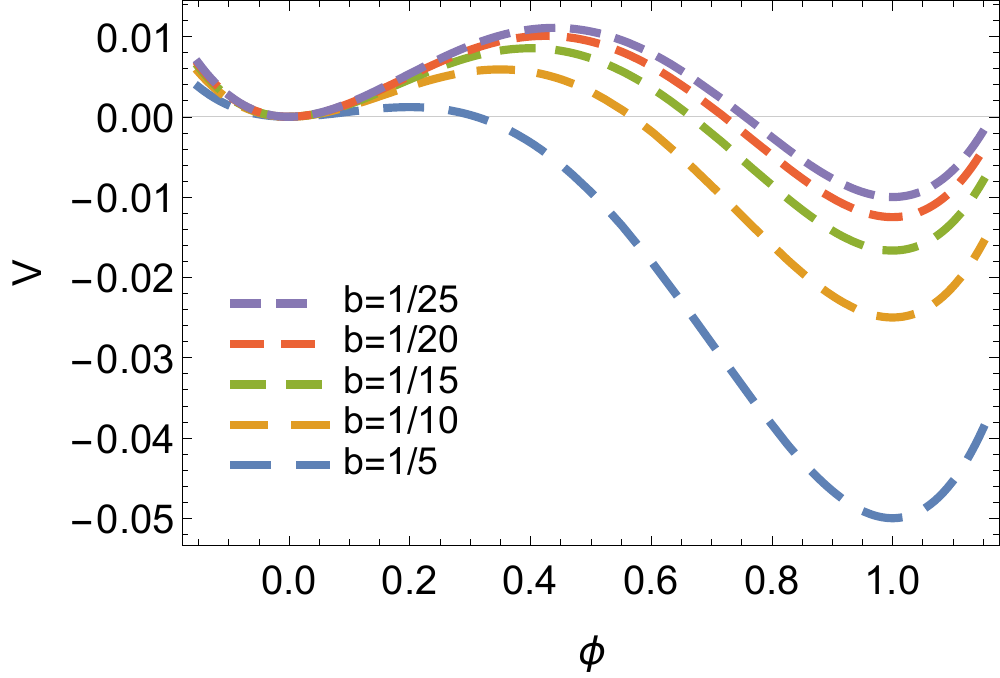}
\caption{
Our toy model potential for different values of $b$ parameter. In this example vacuum energy vanishes $c=0$. Different choices of vacuum energy, we will discuss, simply mean adding a constant to the potential.   
\label{fig:potplot}}
\end{center}
\end{figure}

Figure~\ref{fig:potplot} depicts our potential in the range of parameters used throughout the paper.  We fixed the parameter $a=1$ and used different values of $b$ parameter that controls the degeneration of the vacua.  In this example vacuum energy vanishes $c=0$, different choices of vacuum energy, we will discuss later, simply correspond to adding a constant to the potential.    
 %%%%%%%%%%%%%%%%%%%%%%%%%%%%%%%%%%%%%%%%%%%%%%%%%%%%%%%%
%%%%%%%%%%%%%%%%%%%%%%%%%%%%%%%%%%%%%%%%%%%%%%%%%%%%%%%
\section{Tunneling}\label{sec:tunneling}
%%%%%%%%%%%%%%%%%%%%%%%%%%%%%%%%%%%%%%%%%%%%%%%%%%%%%%%
%\textcolor{magenta}{Whenever the vacuum of our theory is just a local minimum and a new deeper minimum exists in the potential, it is crucial to address the problem of vacuum stability. Since we assume the field begins in a homogeneous configuration in the false vacuum, it is classically unable to cross the barrier between the vacua. However, it is still necessary to address the issue of crossing the barrier due to quantum fluctuations and also the impact of the gravity on the whole process.}

%\red{Only after taking nonperturbative quantum effects into account, the energetically less favorable homogeneous configuration $\phi_{\mathrm{f}}$ is seen to be unstable and have a nonzero chance of instantenously becoming a different, possibly nonhomegenous configuration \cite{Coleman:1980aw}. Inverse of the space-integrated four-density of the tunneling probability is called the vacuum's lifetime.}

Our discussion is based on the standard formalism of Coleman and De Luccia (CDL) \cite{Coleman:1980aw}, which assumes that vacuum decay proceeds through nucleation of true vacuum bubbles within our false vacuum. Notably though, we keep the coupling $\xi$ arbitrary. We will begin by developing a thin-wall approximation \cite{Coleman:1977py,Coleman:1980aw} aimed to include the effects of the non-minimal coupling.
Next we will discuss an even simpler approach assuming that the whole spacetime volume transitions simultaneously \cite{Hawking:1981fz}.
Finally we will describe our exact numerical calculation and use it to discuss the validity of the approximate methods.

The decay probability of the vacuum via bubble nucleation is given by \cite{Coleman:1977py,Callan:1977pt}
\begin{equation}\label{eq:tau}
\Gamma =  A e^{-S},
\end{equation} 
where $S$ is in general the difference of the action integral between final and initial field configurations. Presently, these are respectively the Coleman-DeLuccia bounce $\phi_{\mathrm{CDL}}$ and $\phi_{\mathrm{f}}$ (and we denote $S$ by $S_{\mathrm{CDL}}$) %\textcolor{magenta}{Action of a CDL solution is the difference between the bounce solution and the homogeneous false vacuum background solution}
 \begin{equation}\label{eq:CDLaction}
S_{\rm CDL}=S[\phi_{\rm CDL}]-S[\phi_\f]\;.
\end{equation}
 The prefactor $A$ is derived from quantum corrections to the bounce solution and we do not discuss it in the present paper. 
Therefore to obtain the value of the decay probability we have to calculate the CDL solution and $S[\phi_{\textrm{CDL}}]$.

We are interested in an $O(4)$ symmetric scalar field configuration, $\phi=\phi(\tau)$, with the metric given by $ds^2=d\tau^2 + \rho(\tau)^2(d\Omega)^2$. Here $d\Omega$ denotes an infinitesimal element of the $3D$ sphere and $\rho(\tau)$ is the radius of that sphere. The resulting metric tensor is of the form of the $FRW$ metric with the curvature parameter $k=+1$.
Euclidean action takes the form
\begin{equation}\label{eqn:euclideanaction}
\begin{split}
S_E & =2\pi^2 \int d \tau \rho^3 \left( \frac{1}{2} \dot{\phi}^2+V-\frac{1}{2} \frac{R}{\kappa}\left(1 -\kappa \xi  \phi^2  \right) \right) \\
& = 2\pi^2 \int d \tau  \rho^3 \left( \frac{1}{2} \dot{\phi}^2+V \right)+\frac{3}{\kappa}\left( 1-\xi \kappa \phi^2 \right)
\left(\ddot{\rho} \rho^2+\dot{\rho}^2 \rho-\rho \right)  \\
& = 2\pi^2 \int d \tau \left[ \rho^3 \left( \frac{1}{2} \dot{\phi}^2+V \right)-\frac{3}{\kappa}\left( 1-\xi \kappa \phi^2 \right)
\rho \left(\dot{\rho}^2+1 \right)+6\xi \dot{\phi} \phi \dot{\rho} \rho^2 \right] 
+
\left. \frac{6\pi}{\kappa}\left( 1 - \kappa \xi \phi^2 \right) \rho^2 \dot{\rho} \right|_{0}^{\tau_{\rm max}} \, 
\end{split}
\end{equation}
where $\dot{\phi} = \frac{d\phi}{d\tau}$ and $R=-6\left( \frac{\ddot{\rho}}{\rho}+ \frac{\dot{\rho}^2}{\rho^2} -\frac{1}{\rho^2} \right)$. We integrated by parts the Lagrangian density to get rid of the term proportional to $\ddot{\rho}$, thus acquiring the last boundary term instead. In case of dS false vacuum the boundary term always vanishes as we will see in the next Section.

From the above action \eqref{eqn:euclideanaction} we obtain the equation of motion of the scalar field,
\begin{equation}\label{eq:EFEOM}
 \ddot{\phi}+3\frac{\dot{\rho}}{\rho} \dot{\phi}
-\xi \phi R
 =\frac{\partial V}{\partial \phi}\, ,
\end{equation}
the second Friedman equation,
\begin{equation}\label{eqn:Fried2}
\ddot{\rho}= \frac{\kappa \rho}{ 3\left(1-\kappa \xi \phi^2\right)} 
\left(-\dot{\phi}^2-V+3\xi \left(\dot{\phi}^2+\ddot{\phi}\phi+\dot{\phi}\phi\frac{\dot{\rho}}{\rho} \right)
\right)\, ,
\end{equation}
and the first Friedman equation
\begin{equation}\label{eqn:Fried1}
\dot{\rho}^2=
1+ \frac{\kappa \rho^2}{3(1-\kappa \xi \phi^2)}\left(\frac{1}{2}\dot{\phi}^2-V + 6 \xi \dot{\phi}\phi\frac{\dot{\rho}}{\rho}
\right)\, .
\end{equation}
Using this last equation we can also further simplify the action \eqref{eqn:euclideanaction} to get rid of term proportional to $\dot{\rho}$,
\begin{equation}\label{eqn:euclideanactionsimple}
S_E = 4\pi^2 \int d \tau \left[ \rho^3 V -\frac{3\rho}{ \kappa}\left( 1-\xi \kappa \phi^2 \right) \right] +\left. \frac{6\pi}{\kappa}\left( 1 - \kappa \xi \phi^2 \right) \rho^2 \dot{\rho} \right|_{0}^{\tau_{\rm max}}.
\end{equation}

One can show that scale factor $\rho$ crosses zero at least once \cite{Guth:1982pn}. 
Without loss of generality we chose value of $\tau$ of the first zero to be $\tau=0$ and the other at $\tau_{\rm max}$.
The appropriate boundary conditions then read
\begin{align}\label{eqn:boundaryconditions}
\dot{\phi}(0)&=\dot{\phi}(\tau_{\rm max})=0 \nonumber \\
\rho(0)& =0 \nonumber \\
 \rho(\tau_{\rm max}) & =0, \quad  \quad \quad  \quad\quad{\rm  (for \ dS \ false \ vacuum)} .
 \nonumber
 \\
  \rho(\tau_{\rm max}) & =\rho_{\rm max}\neq 0 \quad \quad {\rm  (for \ Minkowski \ false \ vacuum)} .
\end{align}
Using the definition of $R$, the smooth behaviour necessary in our calculation is not easy to obtain numerically as the second power of $\rho$ appears in the denominator. 
Thus, it is much more convenient for numerical calculations to express the scalar curvature using the Friedman equations as,
\begin{equation}\label{eqn:R}
R=-6\left( \frac{\ddot{\rho}\rho+\dot{\rho}^2-1}{\rho^2} \right)
=
 \frac{\kappa}{ \left(1-\kappa \xi \phi^2\right)} 
\left(\dot{\phi}^2+4V-6\xi \left(\dot{\phi}^2+\phi\ddot{\phi}
+3\dot{\phi}\phi\frac{\dot{\rho}}{\rho}\right)
\right).
\end{equation}
Now $R$ contains only the Hubble parameter that already appears in the scalar field's EOM and thus has to be numerically stable. 

\begin{figure}[ht]
\begin{center}
\includegraphics[height=4.6cm]{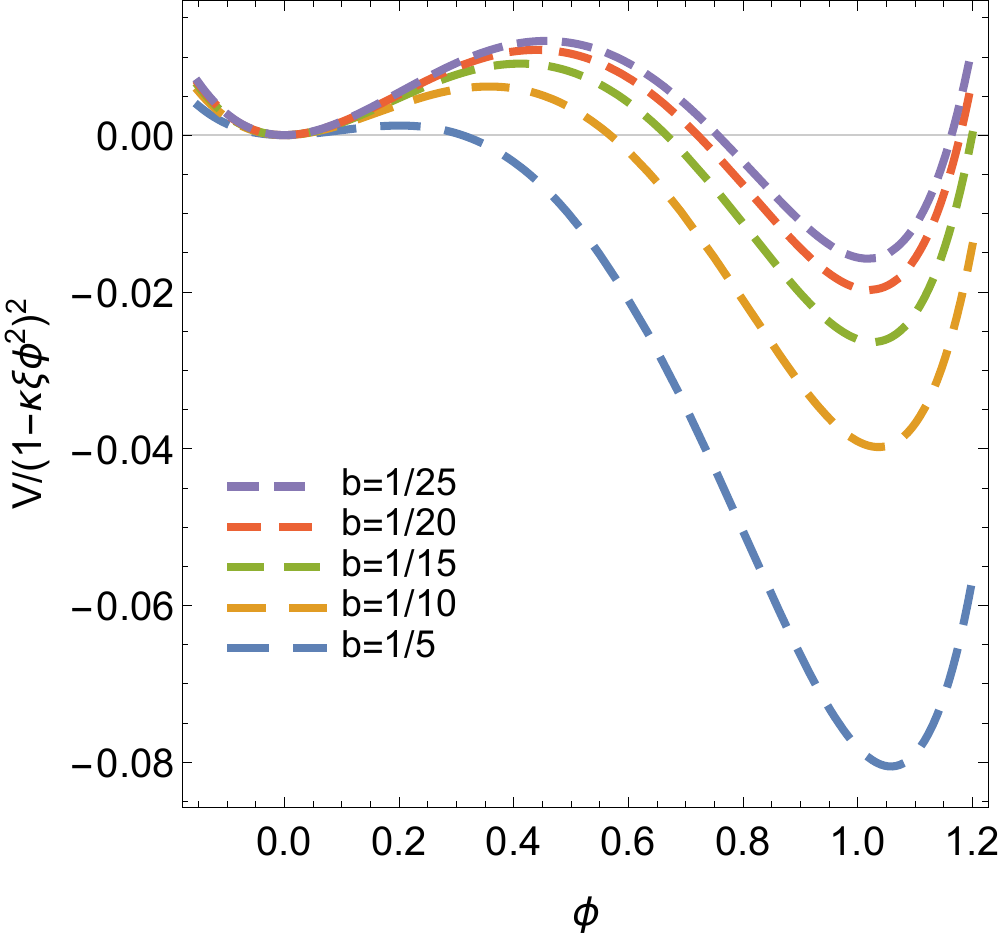}
\includegraphics[height=4.6cm]{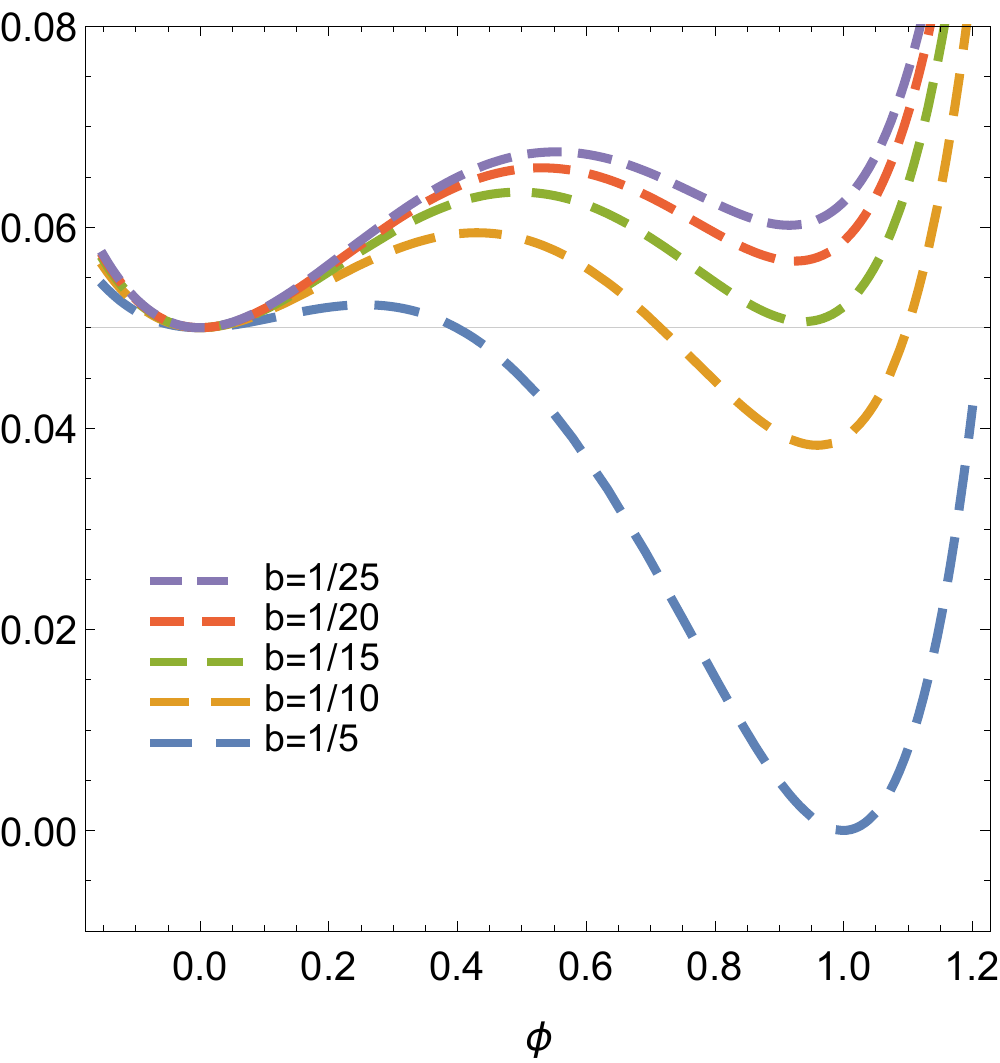}
\includegraphics[height=4.6cm]{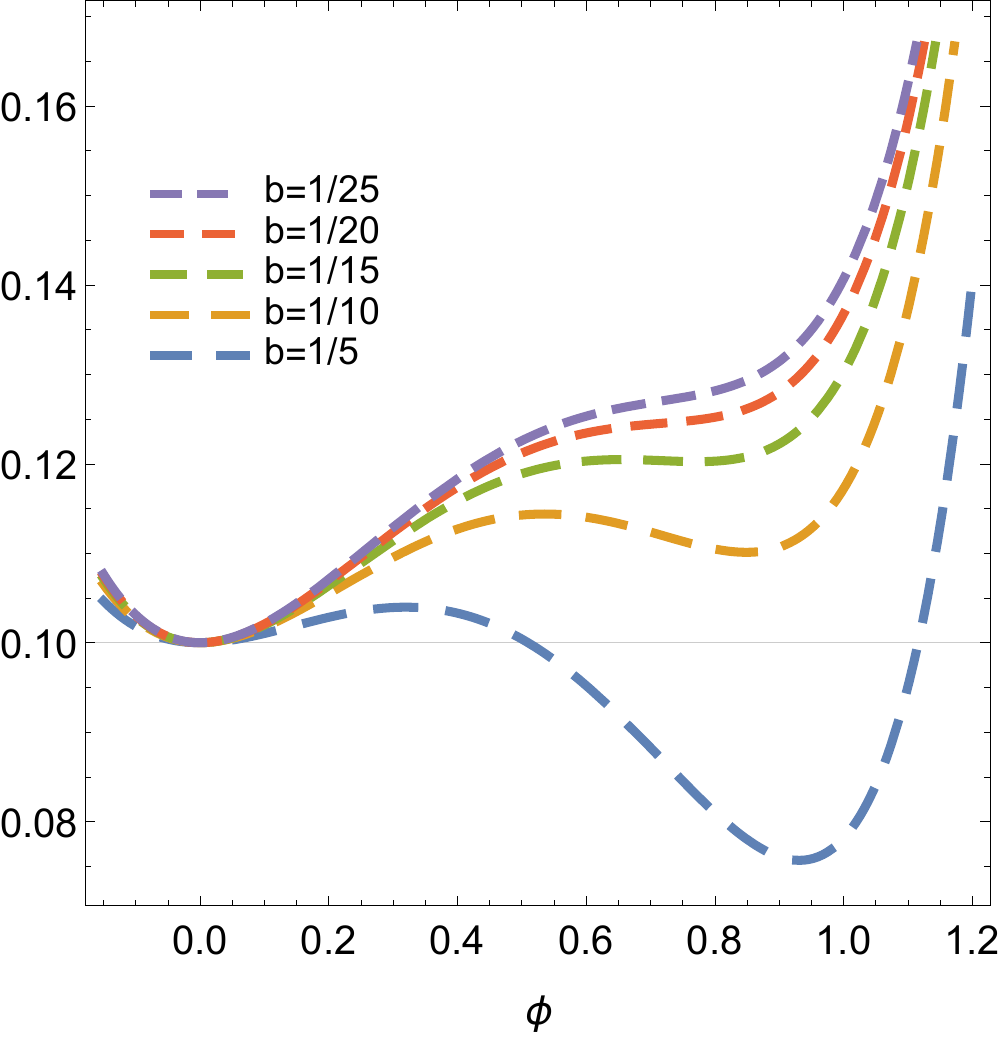}
\caption{
Modified potential $V/(1-\kappa \xi \phi^2)^2$ for different choices of the vacuum energy $c$ and with the non-minimal coupling set to $\xi=0.2$. The value of constant setting the false vacuum energy was set to $c=(0, \, 0.05, \, 0.1)$ from left to right.
\label{fig:modpotplot}}
\end{center}
\end{figure}

In order to calculate gravitational background energy we assume a constant field configuration, with results in the simplified first Friedmann equation \eqref{eqn:Fried1}
\begin{equation}
\frac{d \rho}{d \tau}=\sqrt{1-\frac{\kappa \rho^2 V}{3\left(1-\kappa \xi \phi^2\right)}}\, ,
\end{equation}
where $V=V(\phi)$ and $\phi$ is our chosen constant field value. This allows us to change variables in \eqref{eqn:euclideanactionsimple} and integrating over all space we obtain the action of the background  from \eqref{eq:CDLaction},
\begin{equation}\label{eqn:backgroundaction}
\begin{split}
S[\phi_{\rm f}] & =-\frac{24 \pi^2 (1- \kappa \xi \phi_{\rm f}^2)^2}{\kappa^2 V_{\rm f}}
 \quad \  ({\rm for \ dS}) \\
S[\phi_{\rm f}] & =0  \quad \quad \quad \quad \quad \quad \quad \quad ({\rm for \ Minkowski})\, .
\end{split}
\end{equation}
In our toy potential $\phi_{\rm f}$ is always set to zero so there is no modification of the false vacuum energy. 
However, the same reasoning is applicable to the true vacuum energy.
This already leads to one of the key features induced by the non-minimal coupling. Namely, this modification can increase the energy of our true vacuum beyond that of the false vacuum (in the case when $V(\phi_{\mathrm{t}})>0$) actually making our false vacuum stable.
This is especially visible for large vacuum energies where the true vacuum can disappear altogether as shown in Figure~\ref{fig:modpotplot}. 
In our calculations we always neglect tunnelling in such cases. Even though the bubble profile can sometimes still be calculated, such bubble is not energetically favourable and would not grow after nucleation.
%%%%%%%%%%%%%%%%%%%%%%%%%%%%%%%%%%%%%%%%%%%%%%%%%%%%%%%%
%%%%%%%%%%%%%%%%%%%%%%%%%%%%%%%%%%%%%%%%%%%%%%%%%%%%%%% 
\subsection{Thin-wall approximation}\label{sec:ThinWall}
Now we can proceed to the thin-wall (TW) approximation including gravity. This method, originating from \cite{Coleman:1980aw}, assumes the true vacuum bubble stretches to some $\bar{\rho}$ having a constant value $V_{\rm t}$ and on the outside of the bubble our solution is identical to the false vacuum $V_{\rm f}$. 
The approximate EOM reads
\begin{equation}\label{eq:EFEOM}
 \ddot{\phi}
-\xi \phi R
 =\frac{\partial V}{\partial \phi}\, ,
\end{equation}
where according to our assumptions the scale factor is piecewise constant so that curvature can be approximated by $R=-6\frac{\ddot{\rho}\rho+\dot{\rho}^2-1}{\rho^2}\approx\frac{6}{\rho^2}$.
Integrating (\ref{eq:EFEOM}) once we obtain 
\begin{equation}
\frac{d \phi}{d \tau}=-\sqrt{2(V-V_t)+\xi R \left(\phi^2 -\phi_{\rm t}^2\right)}.
\end{equation} 
Thus the action of the bubble wall reads
\begin{equation}\label{eqn:bubblewallexpansion}
\begin{split}
B_{\rm wall} & = 2 \pi^2 \bar{\rho}^3 \int_{0}^{\tau_{\rm{max}}} \left[2 (V - V_{\rm t}) + \xi R (\phi^2 - \phi_{\rm t}^2)\right] d \tau \\
&\approx 2 \pi^2 \bar{\rho}^3 \int_{\phi_{\rm f}}^{\phi_{\rm t}} \sqrt{2 (V - V_{\rm t}) + \xi \frac{6}{\bar{\rho}^2} (\phi^2 - \phi_{\rm t}^2)} d \phi \\
&\approx 2 \pi^2 \bar{\rho}^3 \int_{\phi_{\rm f}}^{\phi_{\rm t}} 
 \left(\sqrt{2 (V - V_{\rm t})}+\frac{\xi}{\bar{\rho}^2} \frac{3\left(\phi^2-\phi_{\rm t}^2\right)}{\sqrt{2 (V - V_{\rm t})}} \right) d \phi \\
 &\approx 2 \pi^2 \left( \bar{\rho}^3 \int_{\phi_{\rm f}}^{\phi_{\rm t}}
\sqrt{2 (V - V_{\rm t})} d \phi
+
\xi  \bar{\rho}\int_{\phi_{\rm f}}^{\phi_{\rm t}} 
\frac{3\left(\phi^2-\phi_{\rm t}^2\right)}{\sqrt{2 (V - V_{\rm t})}}  d \phi \right) \\
&=2\pi^2 \left( \bar{\rho}^3 S_0+\xi \bar{\rho}S_1 \right),
\end{split}
\end{equation}
where we expanded to the first order in $\xi$. $S_0$ is the usual result we would obtain neglecting gravity and $S_1$ is the linear correction due to the non-minimal coupling.
%%%%%%%%%%%%%%%%%%%%%%%%%%%%%%%%%%%%%%%%%%%%%%%%%%%%
In order to calculate gravitational part of the action we again assume a constant field configuration and as in the previous section perform the integral in \eqref{eqn:euclideanactionsimple}. However this time we integrate only to a given radius  $\rho$ to calculate the action of a bubble, obtaining
\begin{equation}
S_{\rm grav}=2\pi^2 \frac{2}{3} \frac{\left(1-\rho^2 \Lambda V\right)^{3/2} -1 }{\Lambda^2 V } \, ,
\end{equation}
where $\Lambda=\kappa/(1-\kappa \xi \phi^2)$ and $\phi$ is our constant field value.

%%%%%%%%%%%%%%%%%%%%%%%%%%%%%%%%%%%%%%%%%%%%%%%%%%%%%%
Using the above results we combine action of the wall and the difference between true and false vacua gravitational contributions to obtain the final expression for action, which reads
\begin{equation}\label{eqn:thinwallaction}
S_{\rm TW}=2 \pi ^2 \left(\bar{\rho}^3 S_0+\xi  \bar{\rho} S_1
-\frac{2}{3}\frac{\left(\left(1-\bar{\rho}^2 \Lambda_{\rm f} V_\text{f}\right)^{3/2}-1\right)}{\Lambda_{\rm f}^2 V_\text{f}}+\frac{2}{3}\frac{\left(\left(1- \bar{\rho}^2 \Lambda_{\rm t}
   V_\text{t}\right)^{3/2}-1\right)}{\Lambda_{\rm t}^2 V_\text{t}}\right),
\end{equation}
where $\Lambda_{\rm f}=\kappa/(1-\kappa \xi \phi_{\rm f}^2)$ and $\Lambda_{\rm t}=1/(1-\kappa \xi \phi_{\rm t}^2)$ are constant field values. 
In the case of Minkowski background ($V_{\rm f}=0$) the false vacuum gravity action should be replaced with the appropriate limit $S_{\rm grav}\xrightarrow{V \rightarrow 0}-2\bar{\rho}^2 / \Lambda$ which gives 
\begin{equation}\label{eqn:thinwallactionminkowski}
S_{\rm TW}=2 \pi ^2 \left(\bar{\rho}^3 S_0+\xi  \bar{\rho} S_1
+\frac{\bar{\rho}^2}{\Lambda_{\rm f}}
+\frac{2}{3}\frac{ \left(\left(1- \bar{\rho}^2 \Lambda_{\rm t}
   V_\text{t}\right)^{3/2}-1\right)}{ \Lambda_{\rm t}^2 V_\text{t}}\right),
\end{equation}
and analogously for the vanishing energy of the true vacuum.

Differentiating the action with respect to $\bar{\rho}$ and again expanding it to the linear order in $\xi$ we obtain a simple bi-quadratic equation for the size of the bubble $\bar{\rho}$,
{\small
\begin{equation}\label{eqn:twbubblesize}
\begin{split}
   \left[ \left( \frac{1}{\Lambda_{\text{f}}^2}-\frac{1}{\Lambda_{\rm t}^2} \right)^2 -3\xi  S_0
   S_1 \left(\frac{1}{\Lambda_{\rm f}^2}+\frac{1}{\Lambda_{\rm t}^2}\right)\right]+
\bar{\rho}^4 \left[
\frac{9 }{2}S_0^2 \left(\frac{ V_{\rm f}}{ \Lambda_{\rm f}}+\frac{ V_{\rm t}}{
   \Lambda_{\rm t}}\right)+ \left( \frac{V_{\rm t}}{\Lambda_{\rm t}} - \frac{V_{\rm f}}{\Lambda_{\rm f}}\right)^2+\frac{81 S_0^4}{16} \right]+
 \\ +
\bar{\rho}^2 \left[-2\left(\frac{
   V_{\rm f}}{\Lambda_{\rm f}^3}+\frac{V_{\rm t}}{\Lambda_{\rm t}^3}\right)-\frac{9 }{2}S_0^2 \left(\frac{1}{ \Lambda_{\rm f}^2}+\frac{1}{ \Lambda_{\rm t}^2}\right)
 +\frac{2}{\Lambda_{\rm f} \Lambda_{\rm t}}\left(
   \frac{ V_{\rm f}}{ \Lambda_{\rm t}}
   +\frac{V_{\rm t}}{\Lambda_{\rm f}}
   \right)
+3 \xi S_0 S_1 \left( \frac{V_{\rm f}}{\Lambda_{\rm f}}+\frac{V_{\rm t}}{\Lambda_{\rm t}}+\frac{9
  }{4} S_0^2\right)\right]=0 \, .
   \end{split}
\end{equation}
}
Identical equation is obtained from both \eqref{eqn:thinwallaction} and \eqref{eqn:thinwallactionminkowski} after simply using $V_{\rm f}=0$ in \eqref{eqn:twbubblesize}.
To obtain our final approximation for the action we solve the above equation and plug the result back into (\ref{eqn:thinwallaction}) (or \eqref{eqn:thinwallactionminkowski} if the vacuum energy vanishes). For $S_0$ we use the flat space-time relation $S_0=\rho_0 (V_{\rm f}-V_{\rm t})/3$, where $\rho_0$ is the size of the bounce  obtained numerically neglecting gravity (as explained below), while $S_1$ is given by (\ref{eqn:bubblewallexpansion}).
We also checked that expanding to the second order in $\xi$ does not improve our results. In general this correction only slightly increases the action. As we will see later on, this method overestimates the correct result, and so we can say that the error of this approximation comes from our assumption on the shape of the bounce rather than from expanding in the non-minimal coupling $\xi$.

In the absence of gravity our equation of motion for the scalar field simplifies to 
{\begin{equation}\label{eq:FlatEOM}
 \ddot{\phi}+\frac{3}{\tau} \dot{\phi}
 =\frac{\partial V}{\partial \phi}\, .
\end{equation}}
To obtain a finite action we need to satisfy the boundary conditions
\begin{align}
\dot{\phi}(0)&=\dot{\phi}(\tau_{\rm max})=0 \nonumber \\
\lim_{\tau \to \infty} \phi& = V_{\rm f}   \, .
 \nonumber
\end{align}
We solve this equation numerically using the shooting method similar to \cite{Lalak:2014qua}. Next we find the bubble size $\rho_0=\tau \left(\phi=\frac{V_{\rm t}+V_{\rm f}}{2} \right)$ crucial for the bubble tension and use it in \eqref{eqn:thinwallaction}.
We use this numerically obtained bubble size as it is much more accurate than the simple flat spacetimete thin-wall result. Thus we can discuss the validity of thin-wall inclusion of gravity without worrying about the initial flat spacetime error. In what follows we refer to the action of this solution completely neglecting gravity as $S_{\rm flat}$. 

\subsection{Hawking-Moss solution}
Essentially, HM instantons  simply describe the probability for a whole horizon volume to transition to the top of the barrier (and continue by a classical roll-down). 

The action of such an instanton is just the difference between action of our false vacuum and the energy of a homogenous solution on top of the potential barrier. Including the modification of these energies from non minimal coupling as described in 
\eqref{eqn:backgroundaction} we get 
\begin{equation}
S_{\rm HM} =\frac{24 \pi^2 (1- \kappa \xi \phi_{\rm max}^2)^2}{\kappa^2 V_{\rm max}}-
\frac{24 \pi^2 (1- \kappa \xi \phi_{\rm f}^2)^2}{\kappa^2 V_{\rm f}},
\end{equation}
where $\phi_{\rm max}$ and $V_{\rm max}$ correspond potential and field values at the top of the barrier.

\subsection{Numerical calculation of the CDL bounce}
\begin{figure}[ht]
\begin{center}
\includegraphics[width=0.4\textwidth]{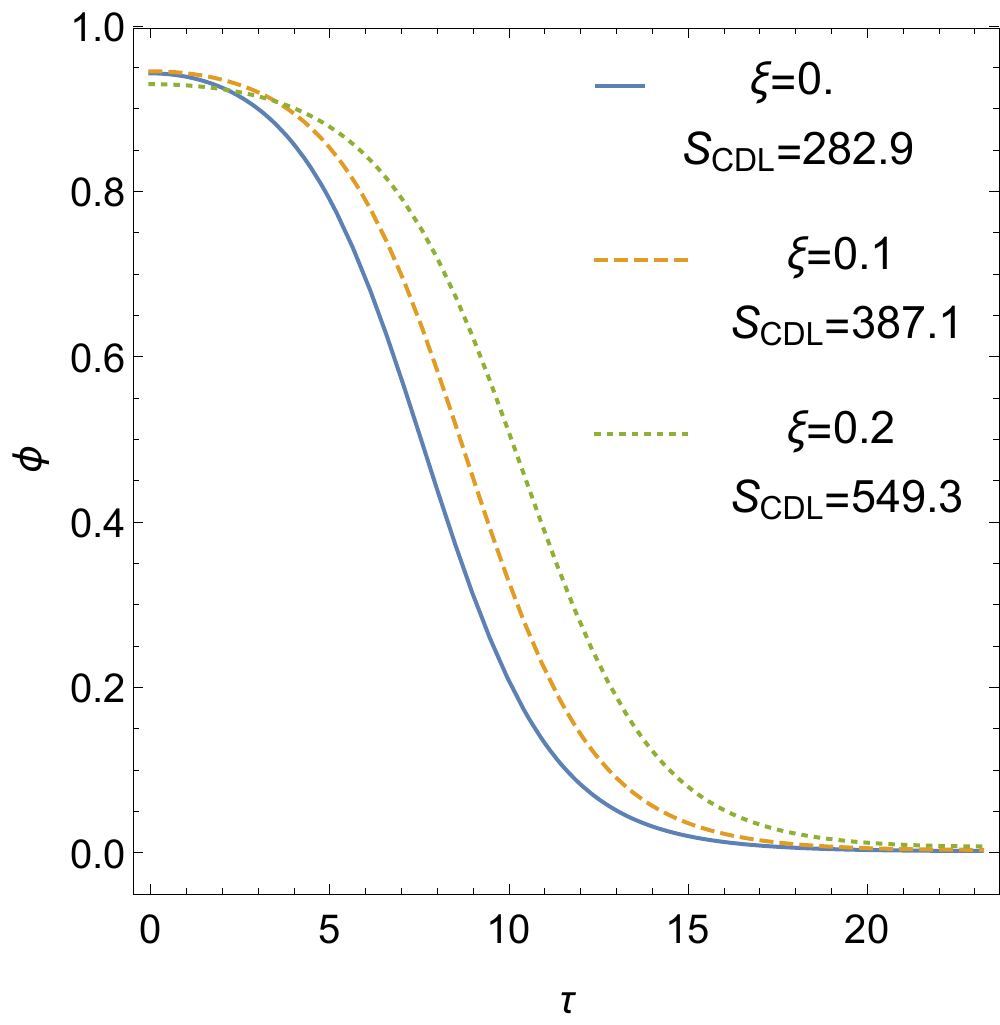}
\includegraphics[width=0.4\textwidth]{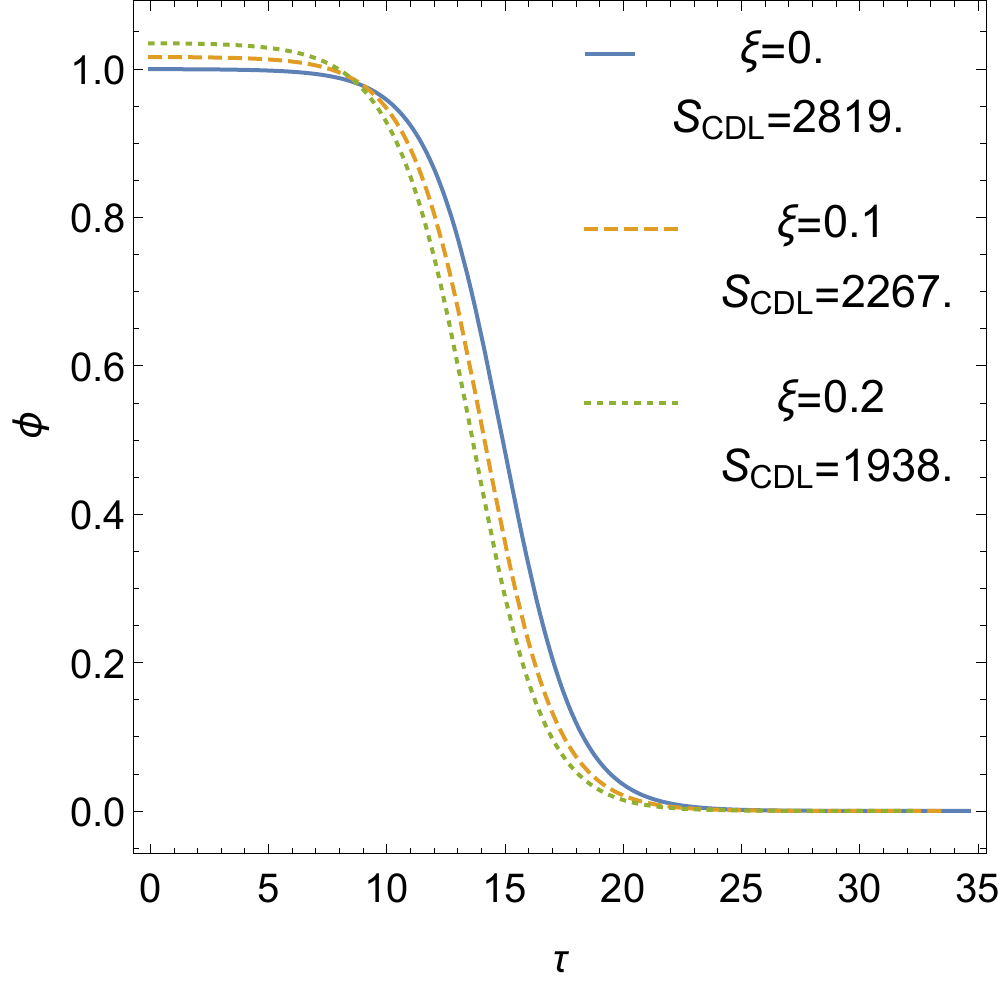}
\includegraphics[width=0.4\textwidth]{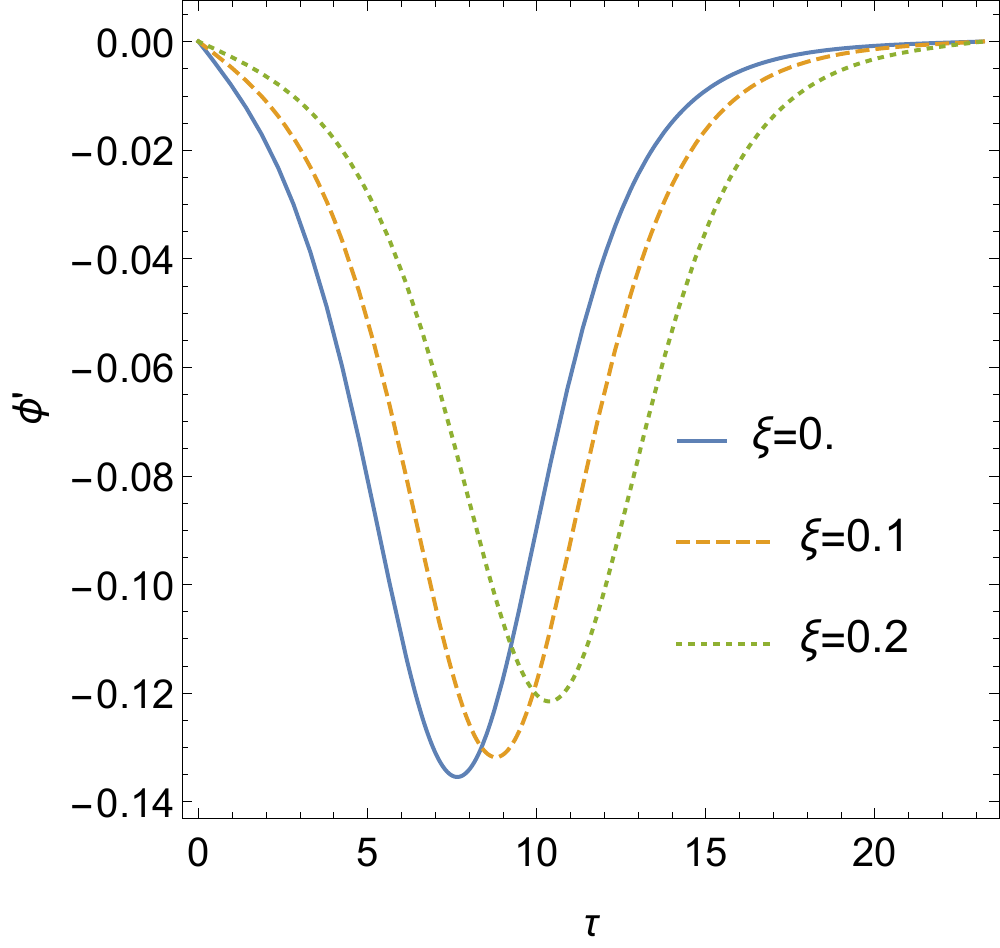}
\includegraphics[width=0.4\textwidth]{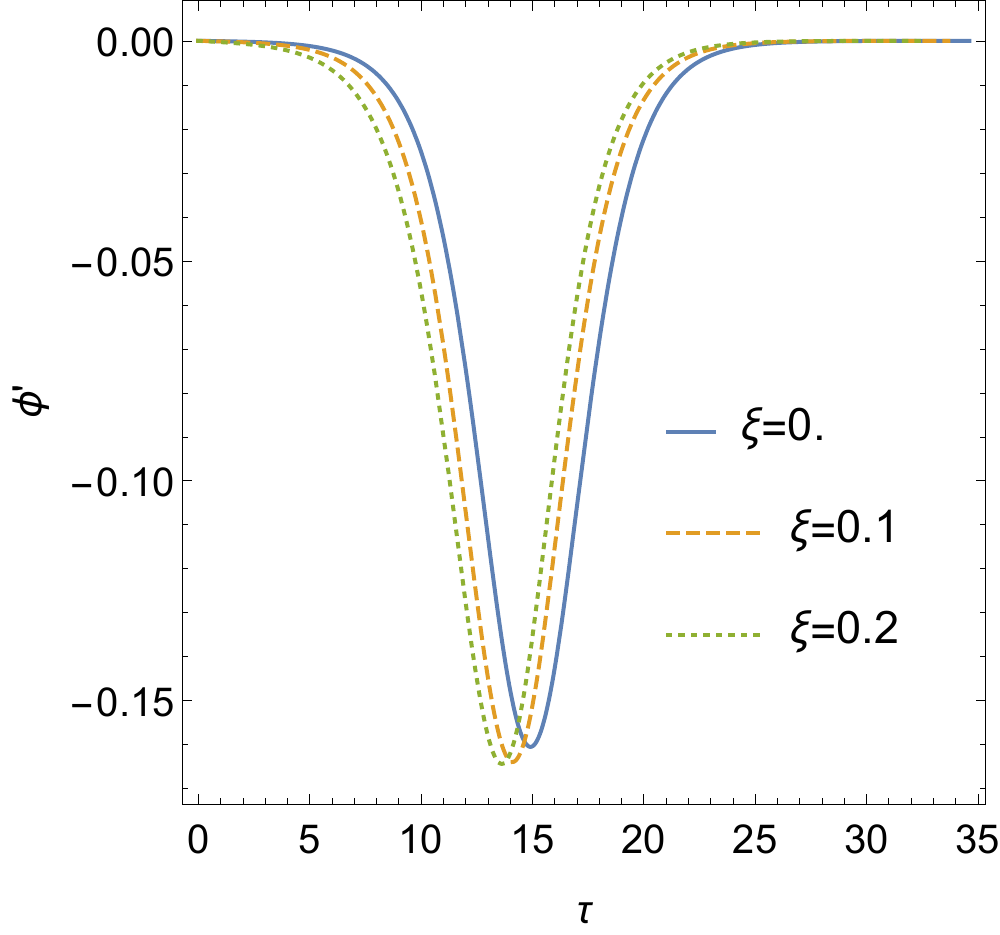}
\includegraphics[width=0.4\textwidth]{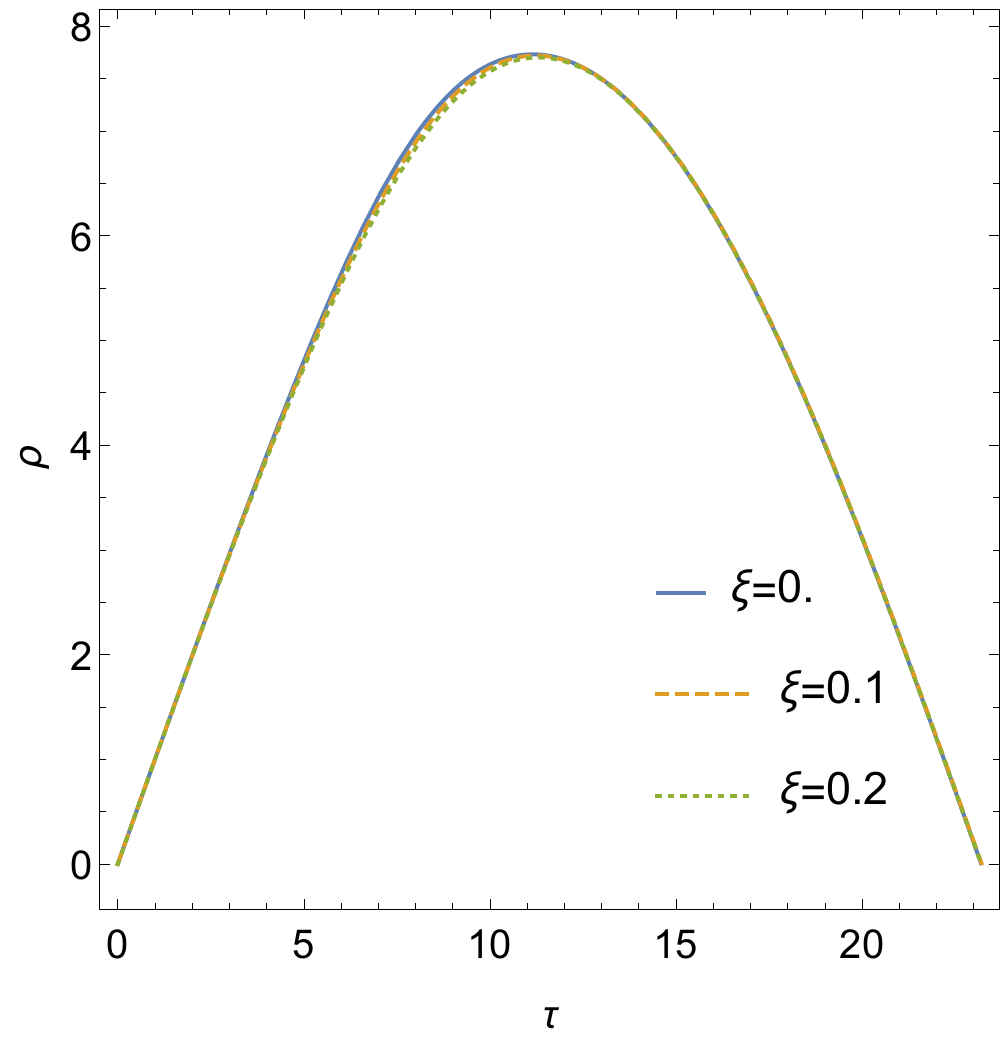}
\includegraphics[width=0.4\textwidth]{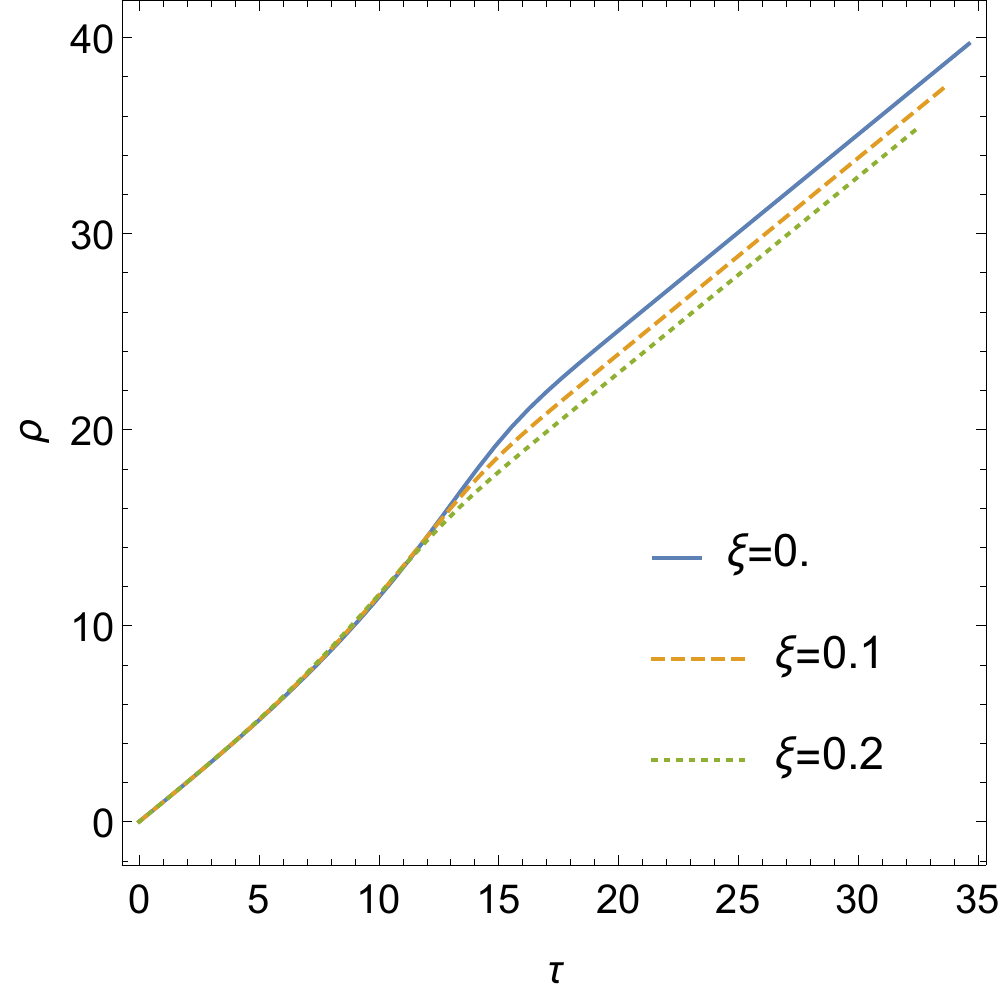}
\caption{
CDL bubble profiles, tunnelling from dS false vacuum $c=0.05$ (left panel) and from Minkowski false vacuum (right Panel) for several values of the non-minimal coupling $\xi$. For this example we set the vacua splitting parameter to $b=1/10$.
\label{fig:instantons}}
\end{center}
\end{figure}

In our numerical procedure we solve the coupled scalar EOM \eqref{eq:EFEOM} with the Ricci scalar expressed through the scalar field \eqref{eqn:R} and the second Friedman equation \eqref{eqn:Fried2}.
As boundary conditions we simply set \eqref{eqn:boundaryconditions}, approximating $\rho(0)$ as proportional to initial $\tau = \epsilon$ and $\dot{\rho}=1$.
The corrections coming from expanding our EOM in a Taylor series give contributions which are higher order in $\epsilon$ and can be neglected as this value can be made arbitrarily small.
The final initial condition needed  for our equations is the field value $\phi_0$. We find the correct value of this parameter corresponding to CDL  by a simple undershoot/overshoot method, known from the flat setup (see e.g. \cite{Lalak:2015usa} for details).
Figure~\ref{fig:instantons} shows the resulting bubble profiles and their modification due to the non-minimal coupling.

It is important here to point out that including the boundary term in the action \eqref{eqn:euclideanactionsimple} is crucial when the false vacuum has a vanishing energy. In this case $\rho$ asymptotes to a linear function instead of crossing zero again at $\tau_{\rm max}$ and the boundary term is sizeable.  

%%%%%%%%%%%%%%%%%%%%%%%%%%%%%%%%%%%%%%%%%%%%%%%%%%%%%%%%
%%%%%%%%%%%%%%%%%%%%%%%%%%%%%%%%%%%%%%%%%%%%%%%%%%%%%%%
\subsection{Comparison of results}
After finding the CDL solution for $\phi(\tau)$ and $\rho(\tau)$ we numerically perform the action integral \eqref{eqn:euclideanactionsimple} which is the final result used in \eqref{eq:CDLaction} together with the background action \eqref{eqn:backgroundaction}. This finally allows us to calculate the action of our solution and consequently to obtain the tunnelling probability.

\begin{figure}[ht]
\begin{center}
\includegraphics[width=0.4\textwidth]{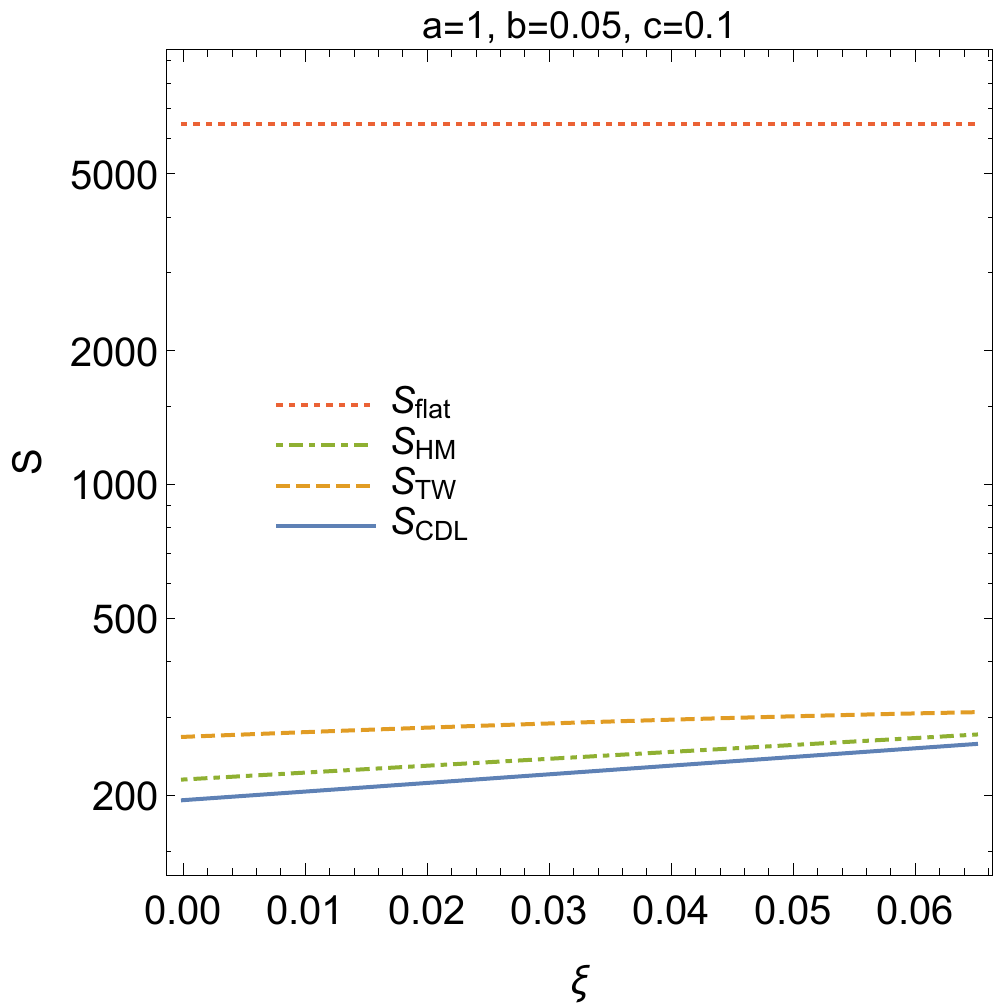}
\includegraphics[width=0.4\textwidth]{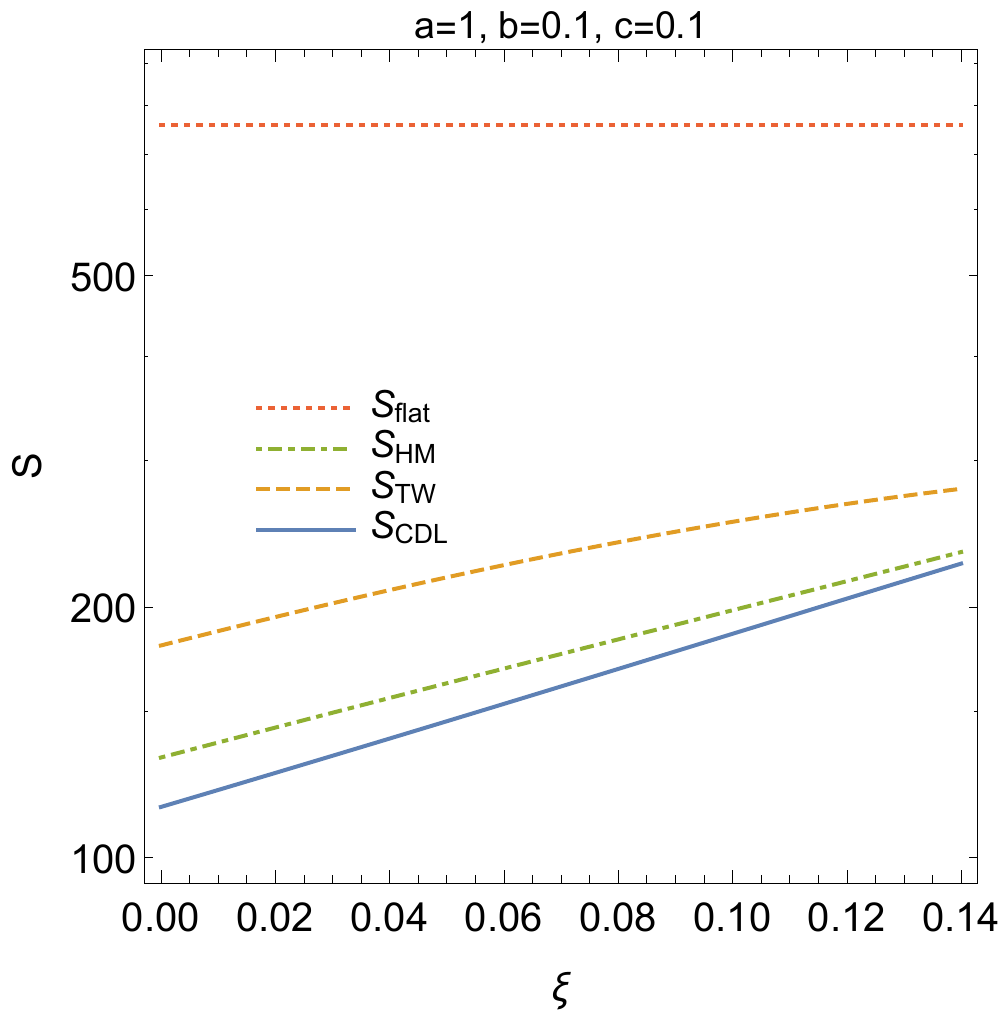}
\includegraphics[width=0.4\textwidth]{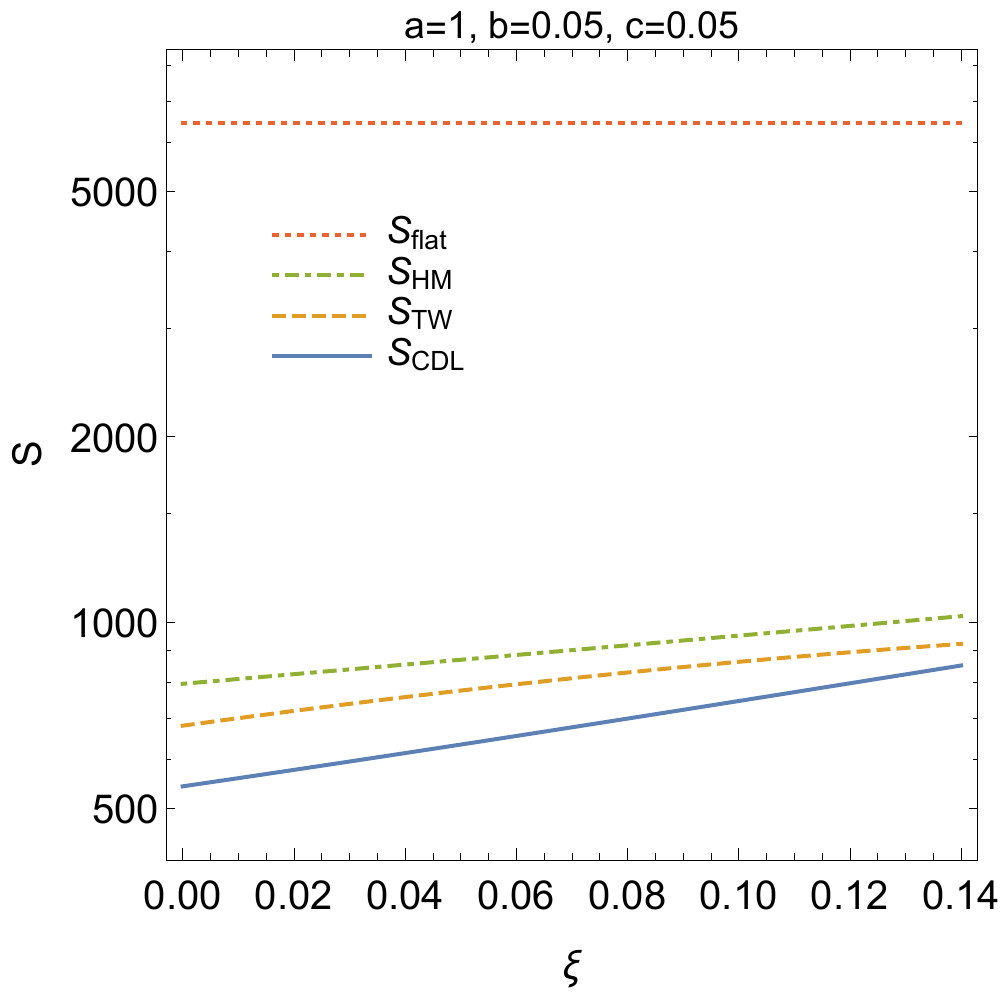}
\includegraphics[width=0.4\textwidth]{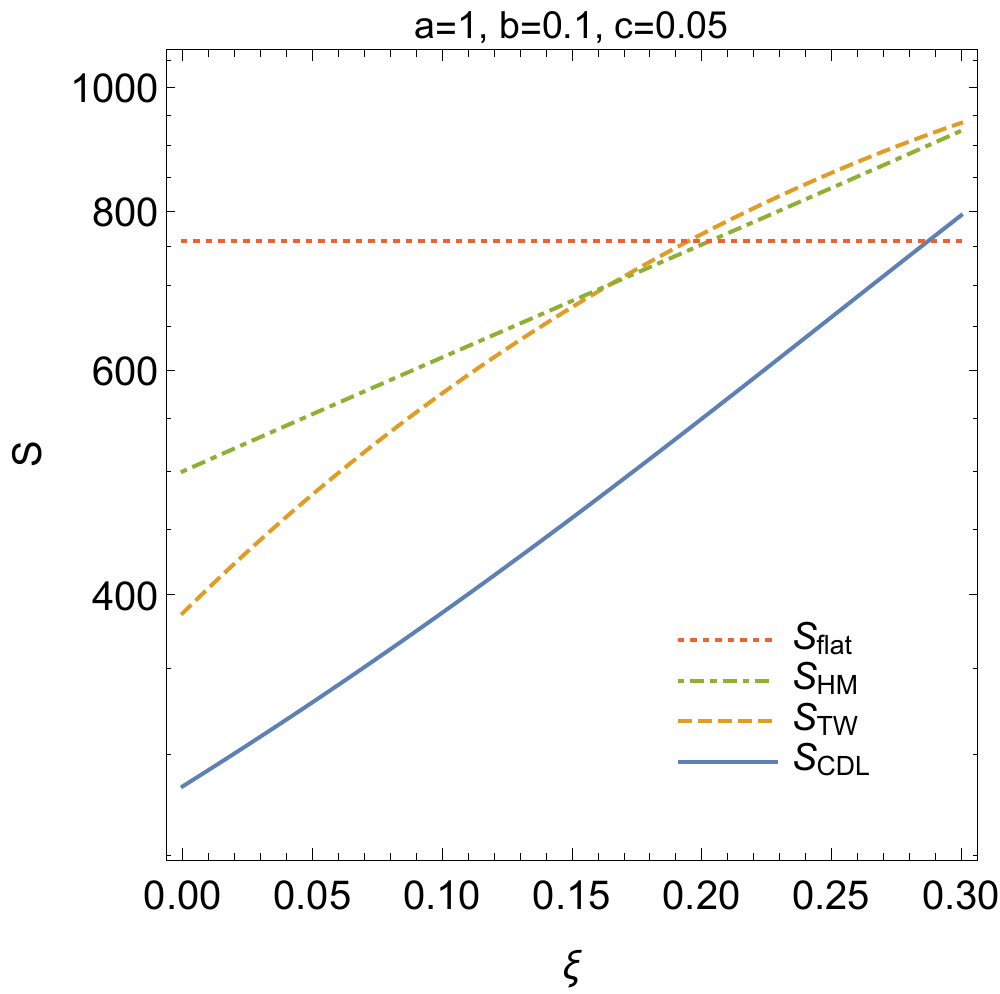}
\includegraphics[width=0.4\textwidth]{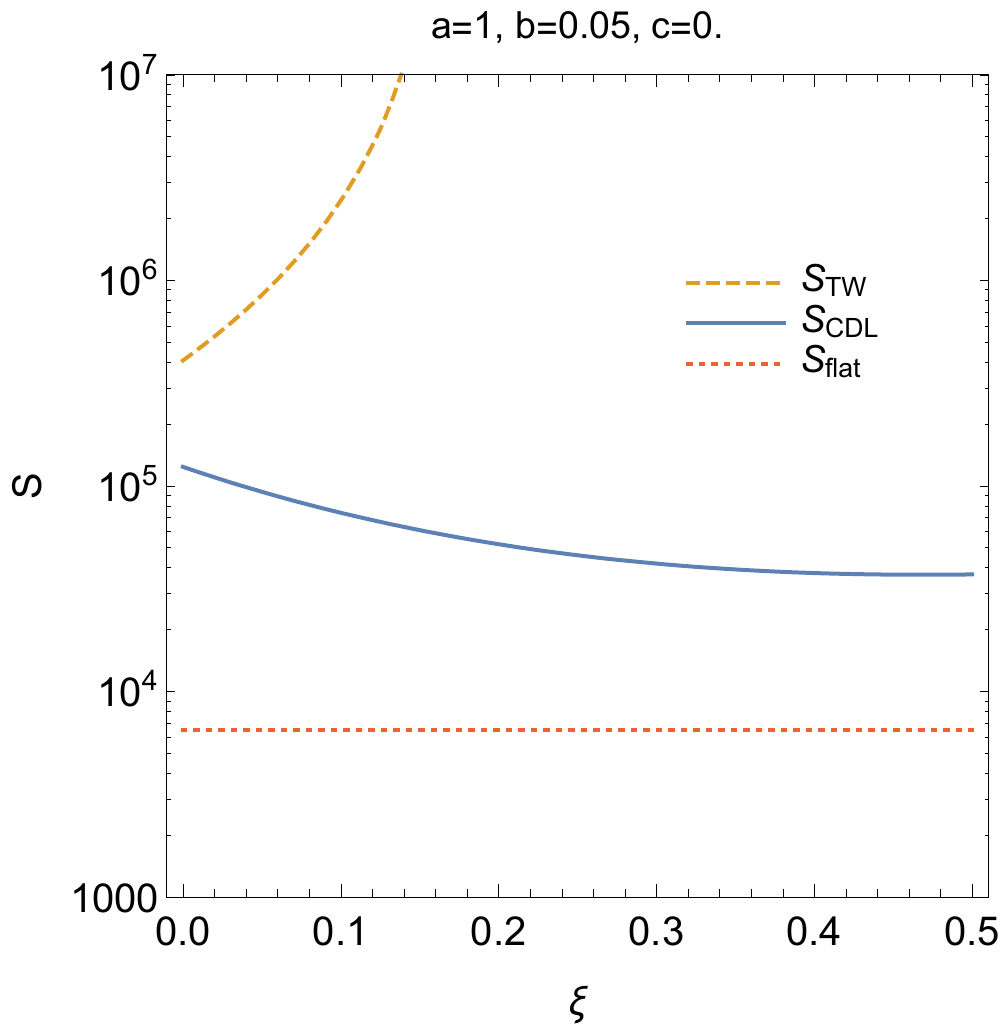}
\includegraphics[width=0.4\textwidth]{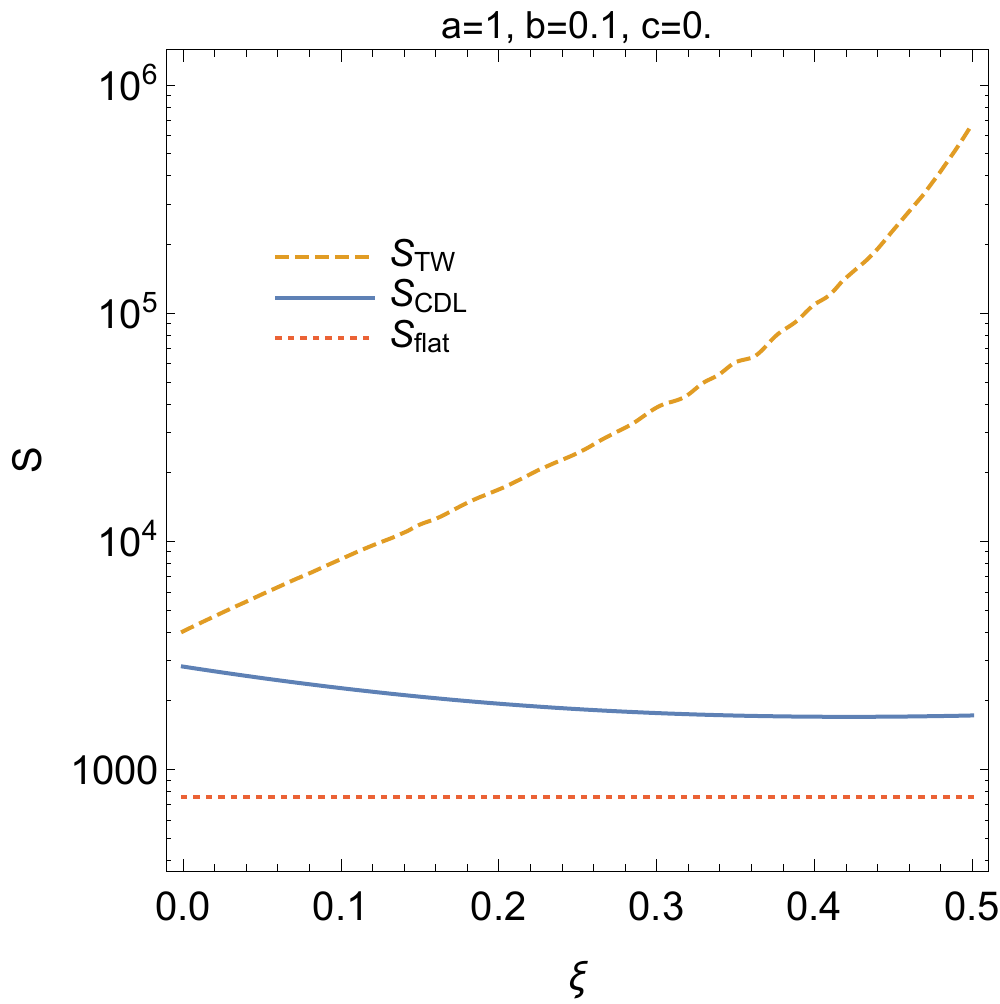}
\caption{
Tunnelling action as a function of non-minimal coupling obtained using four different methods. 
% $S_{\rm CDL}$ is the numerically obtained result fully including gravity,  $S_{\rm TW}$ is the result of our thin-wall approximation, $S_{\rm HM}$ comes from Hawking-Moss solution and $S_{\rm flat}$ is the, numerically obtained, flat spacetime result completely neglecting gravity.
Left column shows results for small parameter $b=0.05$ describing the splitting between vacua, while right column shows results for a bigger value $b=0.1$. Rows show several false vacuum energy densities parametrised by $c=(c=0.1, \ c=0.05, \ c=0)$ from top to bottom. 
\label{fig:action}}
\end{center}
\end{figure}

Figure~\ref{fig:action} shows the logarithm of the resulting action for all methods discussed in this section. $S_{\rm CDL}$ is the numerically obtained result fully including gravity,  $S_{\rm TW}$ is the result of our thin-wall approximation, $S_{\rm HM}$ comes from Hawking-Moss solution and $S_{\rm flat}$ is the, numerically obtained, flat spacetime result completely neglecting gravity. 
%%%%%%%%%%%%%%%%%%%%%%%%%%%%%%%%%%%%%%%%%%%%%%%%%%%%%%%%
%%%%%%%%%%%%%%%%%%%%%%%%%%%%%%%%%%%%%%%%%%%%%%%%%%%

As we can see, both approximations (TW and HM) always overestimate the action.
For relatively large vacuum energies HM solution gives action smaller than thin-wall and is a very good approximation. Considering a smaller vacuum energy, our thin-wall approximation becomes better and the suppression of the action due to gravitational effects lowers.
However both approximations become less accurate as the vacuum energy decreases. This is exemplified in the Minkowski case $c=0$ when the gravitational effects suppress vacuum decay (by increasing the action). Then the HM solution does not exist ($S_{\rm HM}$ would be infinite) and thin-wall severely overestimates the modification due to non zero coupling $\xi$.

\begin{figure}[ht]
\begin{center}
\includegraphics[height=4.7cm]{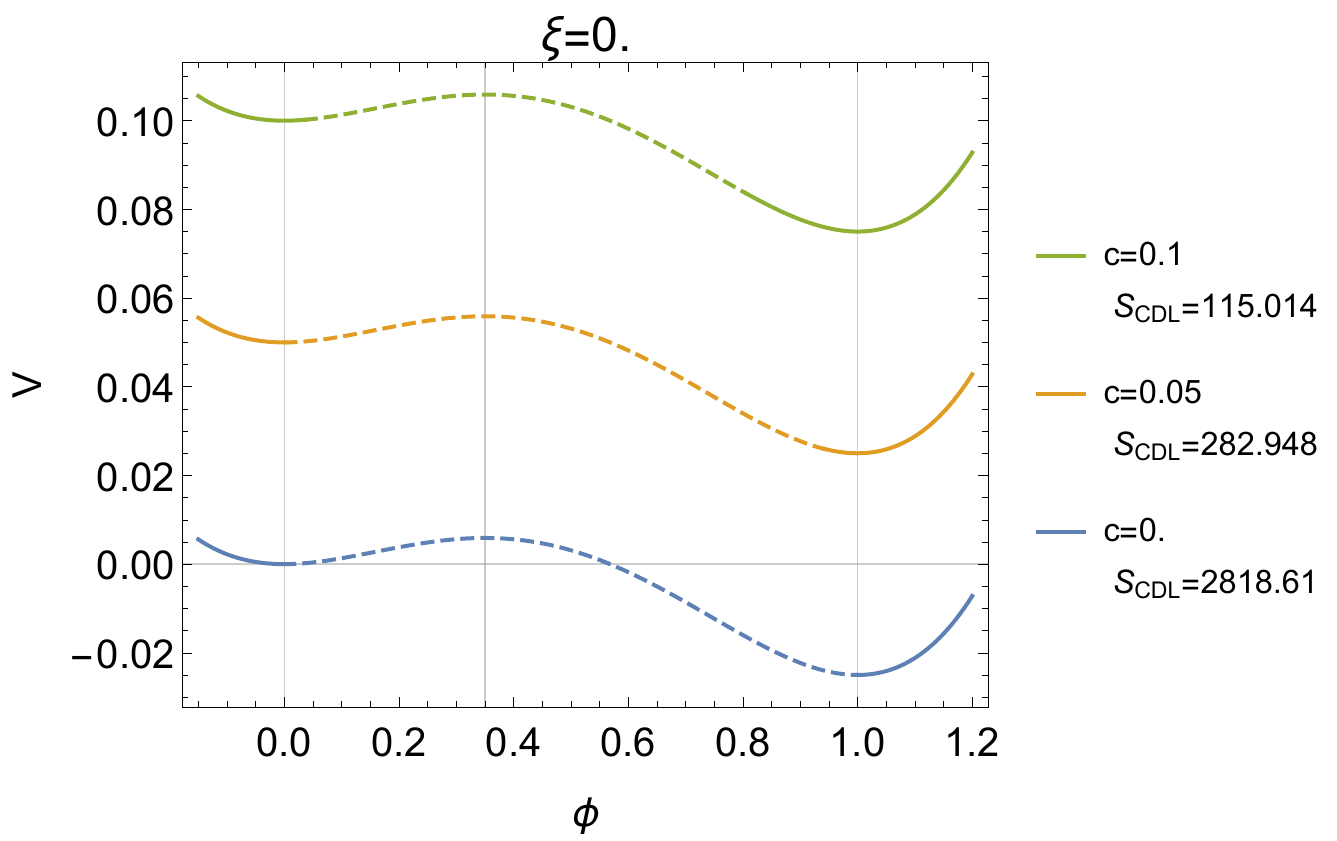}
\includegraphics[height=4.7cm]{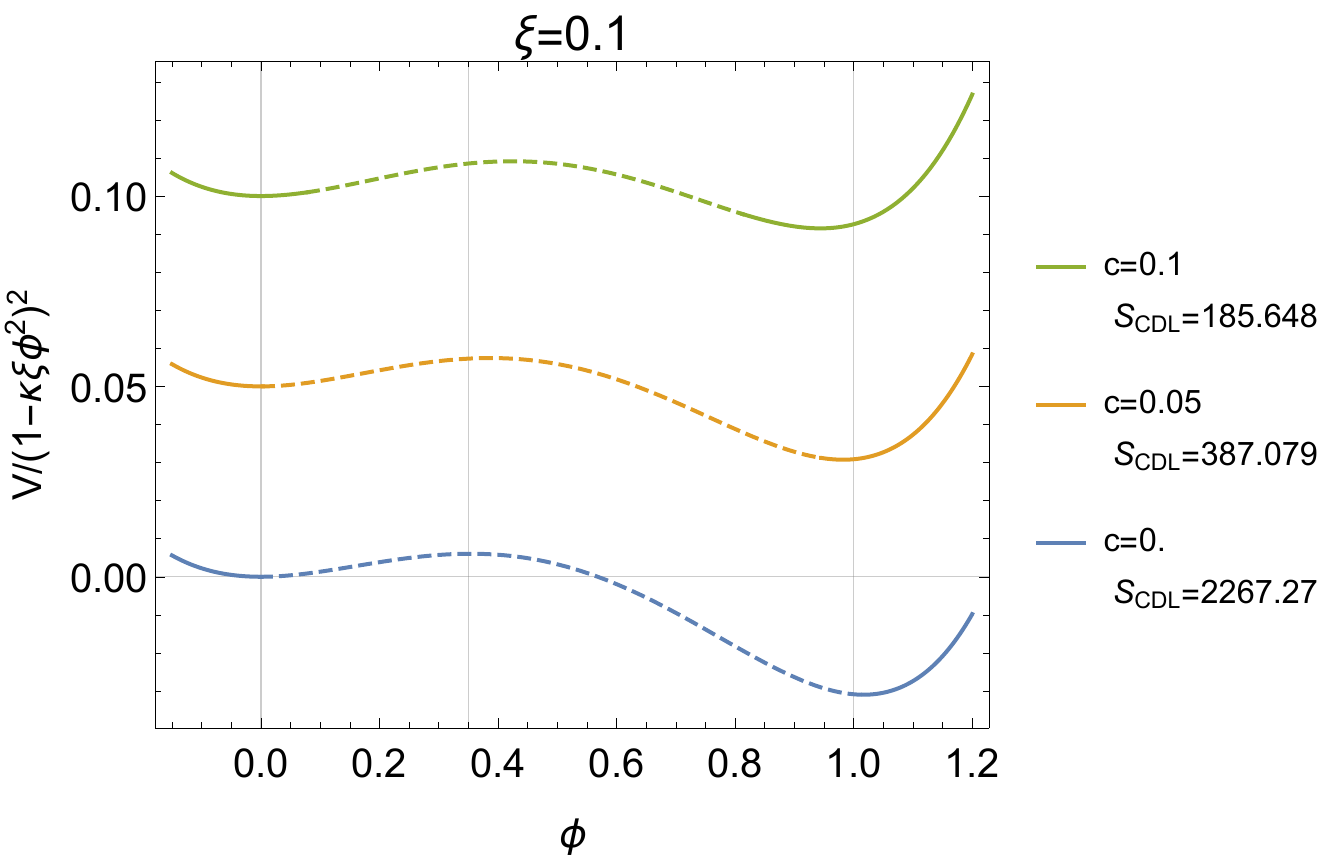}
\caption{
Potentials with different values of the vacuum energy $c$. The part of the potential actually probed by the tunnelling solution is dashed. For this example the non-minimal coupling was set to $\xi=0$ (left panel) and $\xi=0.1$ (right panel) while the vacua splitting parameter $b=1/10$.
\label{fig:thrmaltunneling}}
\end{center}
\end{figure}

We can see that the action quickly decreases as the false vacuum energy increases.
The reason is that in this regime we are essentially dealing with a temperature effect coming from an effective temperature induced by our compact spacetime \cite{Brown:2007sd}.
In this case our bounce solutions do not have to reach the false vacuum but only pass the bubble wall.
We show this in Figure~\ref{fig:thrmaltunneling} which depicts the potentials with different values of the vacuum energy $c$ and part of the potential actually probed by the tunnelling solution.
We also show the same effect in the presence on non-minimal coupling which weakens this effect as it makes the potential more and more flat as the vacuum energy increases, thus also increasing the action. 

As we can see in Figure~\ref{fig:action}, for a fixed positive vacuum energy (given $c$) increasing $\xi$ also results in more flat potential which means the bounce probes only values closer to the top of the barrier making them more similar to the HM solutions. Also when value of $\xi$ is too large the potential becomes too flat and as a result the CDL bounces cease to exist \cite{Hackworth:2004xb,Artymowski:2015mva}. 
Thus, as the vacuum energy decreases larger values of $\xi$ allow tunnelling. 

%%%%%%%%%%%%%%%%%%%%%%%%%%%%%%%%%%%%%%%%%%%%%%%%%%%%%%%%
%%%%%%%%%%%%%%%%%%%%%%%%%%%%%%%%%%%%%%%%%%%%%%%%%%%%%%% 
%%%%%%%%%%%%%%%%%%%%%%%%%%%%%%%%%%%%%%%%%%%%%%%%%%%%%%% 
\section{Conclusions}\label{sec:concl}
In this paper we analysed the vacuum decay process in presence of non minimal coupling to gravity.  
We discuss this issue in a simple model consisting of a single neutral scalar with the generic potential described in Section~\ref{sec:model}.

Section~\ref{sec:tunneling} describe a simple thin-wall solution and provide ready to use formulas needed to compute the decay exponent in a generic model.
We also perform a precise numerical calculation to verify these analytical results.
We show that, while the simple thin-wall approximation would not give a precise result in a specific model, it does provide a correct order of magnitude estimation, especially in the dS false vacuum case, when gravitational correction decreases the stability of the vacuum. 

Our results show that the influence of non-minimal coupling to gravity is very different in cases of Minkowski and dS vacua.
In the latter the decay probability quickly decreases as the coupling grows and in fact the vacuum can be made absolutely stable.
In the flat background case the effect is much weaker and the decay rate increases for small values of the non-minimal coupling.
In this case the thin-wall approximation also works worse, significantly overestimating the increase in action due to non-minimal coupling. 
%%%%%%%%%%%%%%%%%%%%%%%%%%%%%%%%%%%%%%%%%%%%%%%%%%%%%%%%%%%%%%%%%%%%%%%%%%%%%%%%%%%%%%%%%%%%%%%%%%%%
%%%%%%%%%%%%%%%%%%%%%%%%%%%%%%%%%%%%%%%%%%%%%%%%%%%%%%%%%%%%%%%%%%%%%%%%%%%%%%%%%%%%%%%%%%%%%%%%%%%%
\section*{Acknowledgements}
We would like to thank Micha{\l } Artymowski and \L ukasz Nakonieczny for interesting discussions.\\
This work has been supported by the Polish NCN grants DEC-2012/04/A/ST2/00099 and 2014/13/N/ST2/02712, ML was also supported by the doctoral scholarship number 2015/16/T/ST2/00527.
%\appendix
%\section{Some title}
%Please always give a title also for appendices.

%\acknowledgments

%This is the most common positions for acknowledgments. A macro is
%available to maintain the same layout and spelling of the heading.

%\paragraph{Note added.} This is also a good position for notes added
%after the paper has been written.

% The bibliography will probably be heavily edited during typesetting.
% We'll parse it and, using the arxiv number or the journal data, will
% query inspire, trying to verify the data (this will probalby spot
% eventual typos) and retrive the document DOI and eventual errata.
% We however suggest to always provide author, title and journal data:
% in short all the informations that clearly identify a document.

\bibliographystyle{JHEP}
\bibliography{CDLbibliography}

% Please avoid comments such as "For a review'', "For some examples",
% "and references therein" or move them in the text. In general,
% please leave only references in the bibliography and move all
% accessory text in footnotes.

% Also, please have only one work for each \bibitem.

\end{document}